\documentclass[sigconf]{acmart}

\AtBeginDocument{%
  \providecommand\BibTeX{{%
    \normalfont B\kern-0.5em{\scshape i\kern-0.25em b}\kern-0.8em\TeX}}}

\usepackage{graphicx}
\usepackage{threeparttable}
\usepackage{booktabs}
\usepackage{framed}
\usepackage{comment}
\usepackage{hyperref}
\usepackage[export]{adjustbox}
\usepackage[false]{anonymous-acm}

\copyrightyear{2022}
\acmYear{2022}
\setcopyright{acmcopyright}\acmConference[ICSE-SEET '22]{44nd International Conference on Software Engineering: Software Engineering Education and Training}{May 21--29, 2022}{Pittsburgh, PA, USA}
\acmBooktitle{44nd International Conference on Software Engineering: Software Engineering Education and Training (ICSE-SEET '22), May 21--29, 2022, Pittsburgh, PA, USA}
\acmPrice{15.00}
\acmDOI{10.1145/3510456.3514150}
\acmISBN{978-1-4503-9225-9/22/05}

\setcopyright{none}
\renewcommand\footnotetextcopyrightpermission[1]{}
\usepackage{fancyhdr}
\AtBeginDocument{%
    \addtolength{\footskip}{2.0\baselineskip}%
    \fancyfoot[L]{\textit{\textbf{This is a preprint. Please reference this paper as indicated on the first page.}}}%
}

\begin{document}

\title{A longitudinal case study on the effects of an evidence-based software engineering training}

\authoranon{
\author{Sebastián Pizard}
\affiliation{%
  \institution{School of Engineering, Universidad de la República}
  \city{Montevideo}
  \country{Uruguay}}
\email{spizard@fing.edu.uy}

\author{Diego Vallespir}
\affiliation{%
  \institution{School of Engineering, Universidad de la República}
  \city{Montevideo}
  \country{Uruguay}}
\email{dvallesp@fing.edu.uy}

\author{Barbara Kitchenham}
\affiliation{%
  \institution{School of Computing and Mathematics, Keele University}
  \city{Keele}
  \country{UK}}
\email{b.a.kitchenham@keele.ac.uk}

\renewcommand{\shortauthors}{Pizard et al.}
}

\begin{abstract}
\textit{Context:} Evidence-based software engineering (EBSE) can be an effective resource to bridge the gap between academia and industry by balancing research of practical relevance and academic rigor. To achieve this, it seems necessary to investigate EBSE training and its benefits for the practice.
\textit{Objective:} We sought both to develop an EBSE training course for university students and to investigate what effects it has on the attitudes and behaviors of the trainees.
\textit{Method:} We conducted a longitudinal case study to study our EBSE course and its effects. For this, we collect data at the end of each EBSE course (2017, 2018, and 2019), and in two follow-up surveys (one after 7 months of finishing the last course, and a second after 21 months).
\textit{Results:} Our EBSE courses seem to have taught students adequately and consistently. Half of the respondents to the surveys report making use of the new skills from the course. The most-reported effects in both surveys indicated that EBSE concepts increase awareness of the value of research and evidence and EBSE methods improve information gathering skills.
\textit{Conclusions:} As suggested by research in other areas, training appears to play a key role in the adoption of evidence-based practice. Our results indicate that our training method provides an introduction to EBSE suitable for undergraduates. However, we believe it is necessary to continue investigating EBSE training and its impact on software engineering practice.
\end{abstract}


\maketitle

\section{Introduction}\label{sec_intro}

Evidence-based software engineering (EBSE) aims to improve decision-making related to software development and maintenance by integrating the best current evidence of research with practical experience and human values \cite{kitchenham2004}. EBSE has been widely adopted by researchers. For example, Budgen et al. report having found 178 systematic reviews (SRs) published only in the most highly ranked software engineering journals between 2010 and 2015 \citep{budgen2018}. In contrast, little is known about the adoption of EBSE by other stakeholders (e.g., government or industry practitioners). Although lack of EBSE adoption is known to be a problem \citep{cartaxo2016, kitchenham2015, cartaxo2020, hassler2014, dasilva2011}, to our knowledge, there are only three studies that report the application of EBSE in non-academic settings \cite{kasoju2013, cartaxo2018,lewowski2022}. 

Other disciplines where the adoption of evidence-based practice (EBP) is being studied and promoted, have identified the critical relevance of appropriate training. In particular, several systematic reviews placed the lack of knowledge and skills as one of the most commonly reported barriers to adoption \cite{upton2014, scurlock2015, zwolsman2012, sadeghi2014}. 

Motivated by these findings, we have undertaken a research program intended to contribute to improving EBSE adoption by developing and evaluating an EBSE training initiative appropriate for delivery in a university environment and assessing its possible effects on professional practices. 

The initial stage of our research~\citeanon{pizard2021}, involved undertaking a systematic review (SR) aimed at assessing previous EBSE training initiatives, developing an EBSE course using the learning outcome approach aimed at codifying the knowledge and skill required of future EBSE users, and the delivery and evaluation the EBSE course based on students' performance and initial opinions after the course.

The research reported in this paper provides (1) further evaluation of our training method, and (2) an assessment of the impact that the training had on the attitudes and behaviors of students, particularly those related to their working practices. After the first couse, we delivered the EBSE course two more times and used the same initial evaluation process as we did for the first course which was based on surveys, focus groups, and teacher assessments undertaken as part of the training module. Subsequently, to analyze the impact of training, we conducted two surveys of the course participants (one after 7 months of finishing the last course, and a second after 21 months).

The remainder of this paper is structured as follows. In  Section \ref{sec_related} we briefly summarize work related to EBSE training and the effects of EBSE use in industry. Section \ref{sec_proposal} presents our proposal for teaching EBSE and SRs and the context in which it was delivered. In Section \ref{sec_goals}, we specify our research objectives and questions, and describe the research strategy that was used to address them. Section \ref{sec_analysis} presents the data collection and analysis processes. The results are presented and discussed in Section \ref{sec_results}. Finally, we present our concluding remarks in Section \ref{sec_conclu}.

\section{Related work}\label{sec_related}

Our study seeks to evaluate the impact of an EBSE training module on the subsequent work practices of the trainees. So in this section, we provide an overview of  EBSE training research, and also of uses (or proposals for use) of EBSE aimed at improving collaboration between industry and academia.

\subsection{EBSE training}

Unlike other disciplines in which evidence-based practice training is widely studied (see, for example, \citep{kyriakoulis2016, horntvedt2018, lehane2019, larsen2019}), there are not many studies of EBSE training. In a recent SR \citeanon{pizard2021}, we found only 14 investigations of teaching EBSE (reported in 16 articles: \cite{ribeiro2018, lavallee2014, catal2013, castelluccia2013, carver2013, brereton2011, kitchenham2010, oates2009, brereton2009, turner2008, janzen2009, janzen2008, baldassarre2008, rainer2008, rainer2006, jorgensen2005}). The studies reported EBSE training courses with postgraduates and undergraduate students that took place before 2014 and were carried out by universities in seven countries (i.e., Brazil, Canada, Italy, Norway, Turkey, UK \&\ USA). The main purpose of half the studies was related to the teaching of EBSE, while the rest attempted to study either the EBSE process or  attitudes towards EBSE.  
 
All studies included a practical assignment. Typically, it involved participating in the conduct of a secondary study, i.e., an SLR, a limited SLR, or a mapping study. Only three studies focussed on teaching EBSE rather than on the conduct of SRs \citep{rainer2008, rainer2006, jorgensen2005}. None of the studies reported any subsequent evaluation of possible impacts of the training on the professional practice of those trained.

However, three studies report benefits obtained by graduate students from taking EBSE courses \citep{catal2013, kitchenham2010, baldassarre2008}. Table \ref{table_benefits} summarizes their results. Two of them reported having used questionnaires that the students answered after finishing their practical work \citep{kitchenham2010, baldassarre2008}. The reported benefits were acquiring research skills, being aware of the value of the evidence, and learning to search and organize information. However, Catal does not indicate how he obtained his results \citep{catal2013}, so, despite being included in Table \ref{table_benefits}, we do not use them in the discussion of our own results in Section~\ref{sub_sec_discussion}.

\begin{table}[]
\begin{threeparttable}[b]
\caption{Reported benefits of academic EBSE trainings}
\label{table_benefits}
\small
\begin{tabular}{p{6.2cm}p{1.5cm}}
\toprule
Benefit                                                                                       & Reported as results in \\ \midrule
Acquire or improve research skills                                                            & \citep{catal2013, kitchenham2010}                 \\
Become aware of the value of aggregating evidence                                             & \citep{baldassarre2008}                    \\
Learn how to search the literature and organize results                                       & \citep{ kitchenham2010}                     \\ \hline
\textit{Learn how to assess the relevance, validity or quality of the information on a topic} & \citep{catal2013}           \\
\textit{Practice the use of digital libraries}                                                & \citep{catal2013}     \\      
\bottomrule
\end{tabular}
\begin{tablenotes}
\item Catal \citep{catal2013} did not indicate how he obtained these results.
\end{tablenotes}
\end{threeparttable}
\end{table}

\subsection{Effects of EBSE in practice}

Although EBSE explicitly includes within its steps the transfer of knowledge to professional practice, until now more emphasis has been placed on conducting secondary studies than on transferring its results to practice. To our knowledge, there are only three EBSE application reports in non-academic contexts.

First, Kasoju et al. tried to analyze and improve an automotive testing process using among other things EBSE (which included conducting a systematic review) \cite{kasoju2013}. The improvement proposals received positive comments from practitioners, although their implementation is not reported.

Second, Cartaxo et al. reported the conduct of a rapid review (a limited systematic review) in search of practical recommendations that would help an agile development organization with customer collaboration issues \cite{cartaxo2018}. Although the focus was the evaluation of rapid reviews in software engineering, it is also an application of EBSE. Some of the study recommendations were implemented with good results. The results of the rapid review seemed more reliable to the practitioners than the information they usually used, which increased their confidence in their decisions.

Finally, Lewowski \&\ Madeyski reported the conduct of an SR commissioned by a software development company \cite{lewowski2022}, which sought to learn about advances in predicting code smells using AI to improve its platform for automated code review. The SR was an initial stage in a joint research \&\ development project between practitioners and researchers. The use of the results by the company is not discussed in the paper.

Two other initiatives proposed approaches to transfer knowledge from scientific evidence to SE practice using EBSE as a key practice. First, Cartaxo et al. proposed a model that uses rapid reviews and evidence briefings (one-page reports of evidence) \citep{cartaxo2018b}. Second, Badampudi et al. proposed and evaluated a framework for knowledge transfer based on: identification of knowledge (which can be done through secondary studies), transfer to the medium (e.g., using evidence briefings), and contextualization of evidence (for this the authors propose the use of Bayesian synthesis) \citep{badampudi2019a, badampudi2019b}.

Finally, another two recent publications have explicitly suggested or proposed that EBSE may be a valuable resource to bridging the gap between industry and academia.

Devandu et al. investigated the beliefs of Microsoft programmers and how these were related to current empirical evidence \cite{devanbu2016}. To achieve this, they conducted a survey of 564 Microsoft workers and a case study. They reported that practitioners placed more importance on personal experience and the experience of colleagues than on empirical evidence. Therefore, the authors suggest that, given its successful application in medicine, knowledge of EBP might help practitioners to place more trust on verified evidence.

In an opinion article, Le Goues et al. argued that it is possible to improve the use of research in SE, and to obtain better practical benefits, by better organizing and synthesizing its results \cite{legoues2018}. In summary, they suggest the use of EBSE associated with (1) achieving consensus on a formal framework of levels of evidence (i.e., outcomes from secondary and primary studies) and mechanisms to determine the level of confidence in the research results (taking into account types of methods, execution of studies and strength of evidence), (2) clearly identifying the methods and results of studies, (3) encouraging the publication of secondary studies and (4) educating software engineers on how to use the proposed framework.

To summarize, although there are several proposals to use EBSE to better connect academia with industry, we do not know of any that propose EBSE training as the main strategy nor have we found any studies of the impact of EBSE training on professional practice.

\section{EBSE Training Module for university students}\label{sec_proposal}

In our university, \textanon{Universidad de la República}{}, we have a Computer Science (CS) curriculum (a five-year degree comparable to the IEEE/ACM's proposal for the CS undergraduate curriculum \cite{cscurricula2013}). The program comprises 450 credits with certain minimums by areas, e.g. 70 credits in Mathematics. One credit corresponds to fifteen hours of work required by a course for the adequate assimilation of its content, including classroom hours, assisted work, and personal student work. 

In 2017, we introduced a non-mandatory EBSE and SR course. To take our EBSE course, students must have passed the mandatory course on SE. So, students would take this course during the fourth or fifth year of their degree (270 credits approx). In addition, the EBSE course gives students 7 credits upon approval.

We aimed to teach EBSE fundamental concepts and techniques for practical use. After the course, students should understand fundamental EBSE concepts, identify SE issues that may be addressed by using evidence, assess published secondary studies on SE, and participate in the conduct of SR. A limitation of our course was that it emphasized conducting SRs more than EBSE~\citeanon{pizard2021}.

We used learning outcomes (LOs) to guide both the design of the course and the method we use to assess it. LOs specify what students are expected to know, understand, or be able to demonstrate after the course \cite{kennedy2007}. Due to this paper's focus, we do not discuss the LOs in detail, instead, we describe the course, and the results, in terms of topics (which group several LOs). The details of the LOs used for the course can be found in our previous study \citeanon{pizard2021}. However, we present two LOs as an example (they should be read as knowledge or skills that the student will achieve by the end of the course).

\begin{itemize}
    \item Describe the protocol sections of an SLR.
    \item Participate in the selection stage of primary articles for an SLR with multiple reviewers
\end{itemize}

In summary, we specified more than fifty LOs that defining a syllabus that aims to promote the practical application of EBSE, this includes:
\begin{itemize}
    \item Basic aspects of scientific publications (e.g., how to distinguish between scientific papers from other types of publications) and introduction to research in SE (e.g., what research methods are commonly used in SE and for what purpose). These topics are necessary to understand EBSE and it is necessary to include them because many students do not know them when taking our course.
    \item Introduction to the evidence-based paradigm and characteristics of SRs in SE. Based on chapters 1 to 3 of the reference book on EBSE and SRs \cite{kitchenham2015}.
    \item The process of an SR and characteristics of each of its stages. Namely, SR planning, search for primary studies, study selection, study quality assessment, data extraction from primary studies, analysis in mapping studies, qualitative syntheses, reporting of an SR, and knowldege translation and diffusion). Based on chapters 4-10, 12 and 14 of the EBSE book \cite{kitchenham2015}.
\end{itemize}

The course was organized as an alternating introduction of theoretical and practical content and a weekly follow-up of the students' progress in their team assignment (i.e., the conduct of a secondary study). In addition, based on the difficulties and recommendations reported in previous EBSE training studies \cite{lavallee2014, castelluccia2013, oates2009, kitchenham2010, brereton2011, carver2013, jorgensen2005, rainer2008, baldassarre2008}, we chose the following principles to develop the course:
\begin{itemize}
    \item The students' team assignment workload would be limited in some stages of the EBSE process.
    \item Students would be assisted by teachers to choose their review topic. 
    \item Students who needed to perform iterations in some stages of their team assignment would be supported.
\end{itemize}

The course was 14-weeks long and had one non-compulsory on-site class a week that is 3.5-hours long. Table \ref{table_course_timetable} shows the topics of the classes and the team assignments scheduled for each week. During weeks 11-13 students work on completing their team assignments.

\begin{table*}
\small
\caption{Course timetable, including class topics and student assignments.}
\label{table_course_timetable}
\begin{tabular}{cp{5.5cm}p{10.5cm}}
\\ \toprule

Week &	Topics &	Team assignments \\ \midrule
1 & Basic aspects of scientific publications, Introduction to SE research, Evidence-based paradigm, SRs in SE &	Classify the sections of a scientific paper. \\ 
2 &	Planning an SR	&	Establish the purpose and the need for the SR to be performed by each team. Propose and validate the research questions. \\ 
3 &	Searching for primary studies	&	Define the search strategy for the SR. Create and validate the search string. \\ 
4 &	Study selection	 &	Define inclusion and exclusion criteria for the SR. Define and conduct the selection process, obtaining 20-30 primary studies per student. \\ 
5 &	Study quality assessment	&	Define and conduct the quality assessment procedure for the selected primary studies. \\
6 &	Data extraction from the studies &	Define an extraction form and use it to extract data from the primary studies. \\ 
7 &	Mapping study analysis	&	Classify primary studies according to commonly used schemes and schemes relevant to the research questions. \\
8 &	Introduction to data synthesis, Qualitative synthesis & Use qualitative synthesis to answer the research questions, considering the limitations of the review process. \\
9 &	Reporting a systematic review	&	Report the results and the whole process. \\
10 &	Knowledge translation and diffusion	& 	- \\ 
14 &		 \multicolumn{2}{c}{Deadline for team assignment} \\  \bottomrule
\end{tabular}
\end{table*}

As a practical team assignment, students were required to define and conduct guided activities for an SR. Each team chooses the topic of their SR and the research questions, according to their own interests. The teachers guided them in their selection so that the work, both in scope and complexity, could be tackled in the time available. The teams were made up of two or three students who worked together throughout the course and delivered the final report of their SR in week 14. The weekly class covered the theoretical content of a stage of the SR,  which the students then applied to their own SR before the next class.

Most classes had two main parts: a lecture by a teacher and a reserved time for teamwork and questions to teachers. The lectures usually took less than an hour during which the teacher explained the main concepts of the SR stage that were studied that week. Students were asked to pre-read the material, which was made up of chapters from the EBSE book \cite{kitchenham2015}. Subsequently, the teacher presented the weekly assignment task, which consisted of completing the activities of the SR stage previously discussed.

Students could ask questions or present problems they had about the assignment for the current or previous weeks. The two teachers answered questions and were present in the classroom for the entire class. Students were expected to spend four hours weekly outside of the classroom to complete the required assignment. To discuss their problems during the course, students could use a Moodle platform site, where teachers published material and answered questions. The basic format of the course was unchanged through all three years. In 2018, the course content was extended to introduce rapid reviews and evidence briefings, but otherwise unchanged.

\textanon{Sebastián Pizard}{The first author} was not only responsible for the design and evaluation of the course, but he was also lead researcher of the course evaluation and one of the two teachers on each of the three courses. During 2017 and 2018 he was accompanied by \textanon{Fernando Acerenza}{} and in 2019 by \textanon{Cecilia Apa}{}. All three are active researchers with completed master's degrees and eight years (or more) of experience of conducting lectures and tutorials.

\section{Research goals and method}\label{sec_goals}

The overall objective of our research program is to contribute to the understanding of training as a facilitator for the adoption of EBSE.  In this article, we extend the evaluation of our training module reported in our previous study \citeanon{pizard2021} to include data from two more years, and we report the impact of the training on the participants' subsequent working practices. Specifically, our research questions are:

\begin{framed}
\noindent
\begin{itemize}
    \item[RQ1:] Is our EBSE course adequate to train undergraduate students?
    \item[RQ2:] Does our EBSE course have any impact on the working practices of the students?     
\end{itemize}
\end{framed}

We consider that course adequacy (RQ1) is directly related to the successful achievement of the predefined LOs by the students. These LOs (and the teaching methodology used) are intended for students to understand EBSE and apply it by conducting the steps of the SLR process. The achievement of these LOs (as explained below) was assessed through teacher evaluations of student's assignments and the perceptions of the students themselves.

Similarly, we consider as impacts of the EBSE course (RQ2) any change in attitudes and behaviors of students caused by what they learned. The possible impacts were investigated surveying the students, with which the results are associated with their perceptions. Furthermore, we cannot isolate our course from the rest of the curriculum, with which it could be thought that students would exercise some EBSE practices in other courses. From our knowledge of the curriculum, we understand that there are no other courses that include our syllabus and that this should not be a threat to our work.

We treat our study as a \emph{longitudinal case study}, which, according to Yin, seeks to study \textit{`how certain conditions and their underlying processes change over time'} \citep{yin2017}. Fucci et al. argue that longitudinal studies are useful to study a particular event and its impact, for example, the introduction of a practice in a company \citep{fucci2018}. This research method enables studying dynamics of change in complex and changing contexts such as those related to software engineering \citep{mcleod2011}, and, therefore, to investigate the effects of learning EBSE. 

The collection of data in different waves (i.e., separate time periods) allows better understanding of the evolution of the impact of training. In particular, we can study both their initial impressions of EBSE and any subsequent impact on their professional practice. 
\begin{figure*}[htb!]
\begin{center}
\includegraphics[width=16cm]{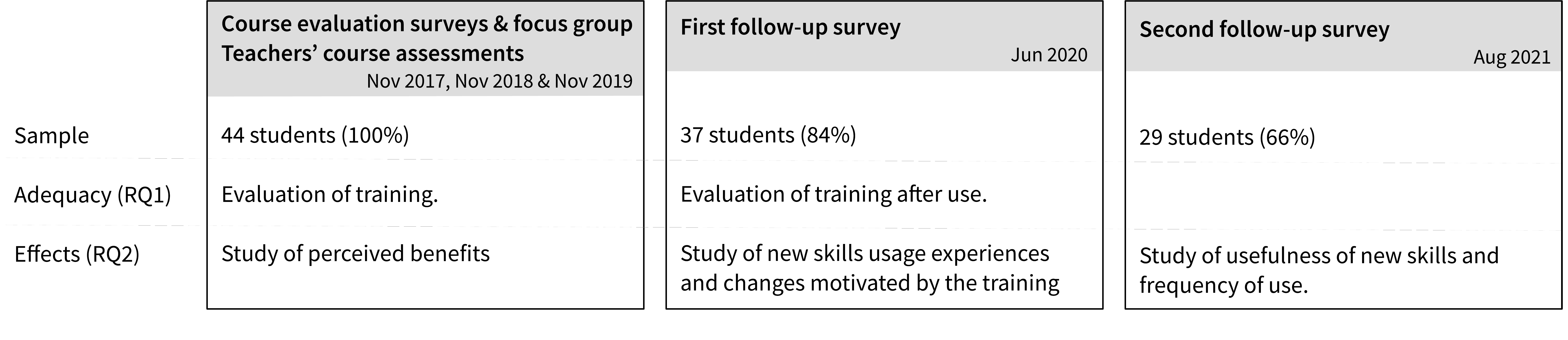}
\end{center}
\caption{Summary of research waves.} \label{fig_stages}
\end{figure*}

The data collection waves carried out are summarized in Figure \ref{fig_stages}. Since it is a longitudinal study, certain aspects of our research design are worth mentioning \citep{fitzmaurice2008}. The amount of time between the waves was chosen for convenience and based only waiting a reasonably long period between the end of the course and circulating our surveys, because we wanted to allow enough time for participants to encounter opportunities to use their training. We sought the participation of all the students of the three courses. Data collection was always anonymous (the tools used allowed checking a single response per subject). Anonymous collection has certain advantages \citep{audette2020}: (1) participants respond more honestly if they trust that they cannot be identified in any way, (2) anonymous data reduces confirmation bias, and (3) anonymous data collection helps to comply with ethical and legal requirements. The most notable disadvantage, in our case, is that we cannot perform intra-individual comparisons across time. We consider that the design used was adequate to meet the research objectives, which are exploratory, while also encouraging participation. The unit of analysis used for most analysis was data from individual students. However, some course assessments were based on the assessment of team projects.

To improve reliability (e.g., to avoid misinterpretation of the data), we used a rigorous pre-planned analysis process. We also used two approaches proposed by Runeson et al. to improve reliability \citep{runeson2012}: systematic tracking of all data, and peer debriefing (i.e., participation of at least two researchers). In addition, to enhance reporting clarity and completeness, we used the O'Brien et al. checklist for reporting qualitative research \cite{obrien2014}.

\section{Participants, Data Collection and Data Analysis}\label{sec_analysis}

This section describes the participants, and presents details about data collection and analysis.

The data collection and analysis procedures can be separated into those performed for the course delivery data and those carried out in the follow-up surveys. All of the surveys were designed and conducted following guidelines and recommendations proposed in the literature \cite{kasunic2005, torchiano2017}.

\subsection{Subjects Selection and Ethical Issues}

Each student was the main unit of analysis, however, on several occasions, it was necessary to analyze student teams, e.g. while assessing their practical assignment.
Our course is non-mandatory and students were encouraged to take it by a typical course information entry on the university website. 

During the first class of the course, the teachers explained to the students that they were researching on EBSE training, and reported the aims and the data collection procedures that they planned to use. They were also told that the information collected was to be managed confidentially (i.e., course assessment reports would not link grades or test scores to individual students or teams), that participation or not in the study would not affect their learning experience or evaluation, and that they could withdraw from the study at any time without leaving the course. For each of the three courses, all students agreed to participate voluntarily and signed an informed consent form. 

When asked to answer each of the two follow-up surveys, the students were also reminded about the objectives of the research, the need for accurate and honest answers, and the confidentiality of any reported information.

\subsection{Courses data collection and analysis}

During each course (2017, 2018, 2019), we collected quantitative and qualitative data. The data include the students' opinions - collected through a survey (will be referred to as SU) and a focus group (FG) - as well as the academic results obtained from the learning assessments (LA). The survey and the focus group were carried out in the last class. The learning assessments include grades achieved in team practical assignments and individual written tests. 

The analysis presented in this paper does not include the individual written tests. In these tests, the results obtained were quite satisfactory in all the courses, although not comparable, since their questions sought to evaluate different learning outcomes.

The details of the data analysis process can be found in our previous study \citeanon{pizard2021}. As significant changes we can report:
\begin{enumerate}
    \item We counted students' responses to whether they had work activity or not and their weekly workload (all obtained through SU). Work positions were analyzed and classified using descriptive coding and the constant comparative method.
    \item Students' opinions (SU) and teachers' assessments (LA) of the achievement of learning outcomes (initially expressed by using a five-point agreement scale) were grouped and counted as positive or negative.
    \item Students' opinions on the benefits of the training (SU) were grouped according to similarity. Opinions that indicated benefits in their work practices were analyzed using descriptive coding and the constant comparative method.
\end{enumerate}

\subsection{First follow-up opinion survey}

Seven months after finishing the last course we conducted the first follow-up survey. The main purpose of the survey was to collect the former students' opinions about any impact the EBSE course had on their work and academic practices. It is worth mentioning that at that time we wanted to know any impact of the course, although later in the following survey we asked only for effects on professional practices.

The three authors defined the purpose of the post-course survey and the initial draft of the survey. Subsequently, \textanon{Pizard}{the first author} created the questionnaire, which was reviewed by \textanon{Kitchenham and Vallespir}{the third and the second authors}. The self-administered questionnaire had six questions. We used a set of closed and open questions to allow participants to explain their answers more completely. The questionnaire was designed in SurveyMonkey and was available from June 23 to July 2, 2020.

The translation of the survey questions (originally in Spanish) is reproduced below, the flow of the survey is presented within square brackets.

\begin{framed}
\small
\noindent
\begin{itemize}
    \item[F1] What year did you take our EBSE course?
    \begin{itemize}
        \item 2017 / 2018 / 2019
    \end{itemize}
    \item[F2] After finishing the course, have you used any of what you have learned in your professional or academic practice? For example, some of the following activities.
    \begin{itemize}
        \item Conducting a secondary study (systematic review, mapping study, or rapid review).
        \item Using an existing secondary study (systematic review, mapping study, or rapid review).
        \item Using search engines for scientific articles.
        \item Undertaking a critical appraisal of reports that compare different technologies.
        \item Other activities, indicate which.
    \end{itemize}       
    \item[F3] \textit{[If at least one option selected in F2]} Do you think the course adequately trained you for those activities? Please explain your answer in detail.
    \begin{itemize}
        \item Yes / No / Comments
    \end{itemize}    
    \item[F4] \textit{[If empty answer to F2]} Do you think that your knowledge about EBSE could be used in the future? For what?
    \item[F5] Did knowing about EBSE MOTIVATE you to improve or change any activity in your professional practice related to software development?
    \begin{itemize}
        \item Yes / No
    \end{itemize}
    \item[F6] \textit{[If answer to F5 was 'Yes']} In which activities, tasks, problems or situations do you perceive that you have changed after learning about EBSE?
\end{itemize}
\end{framed}

To analyze F1, F2, F3, and F5 we counted the number of responses in each category. In F3, F4, and F6 the textual responses were collated and we used descriptive coding and constant comparative method to classify the responses. 

Initially, \textanon{Pizard}{the first author} performed all of the analysis described above. Then, \textanon{Vallespir}{the second author} verified the result of the quantitative responses, reviewed the coding of the F1, F2, and F3, and did a separate analysis of F6. Subsequently, in two meetings they both reached an agreement on the coding of all the questions. Finally, \textanon{Pizard}{the first author} made the grouping of the final codes and description of the results.

\begin{figure}[htb!]
\centering
\fbox{\includegraphics[width=8.2cm]{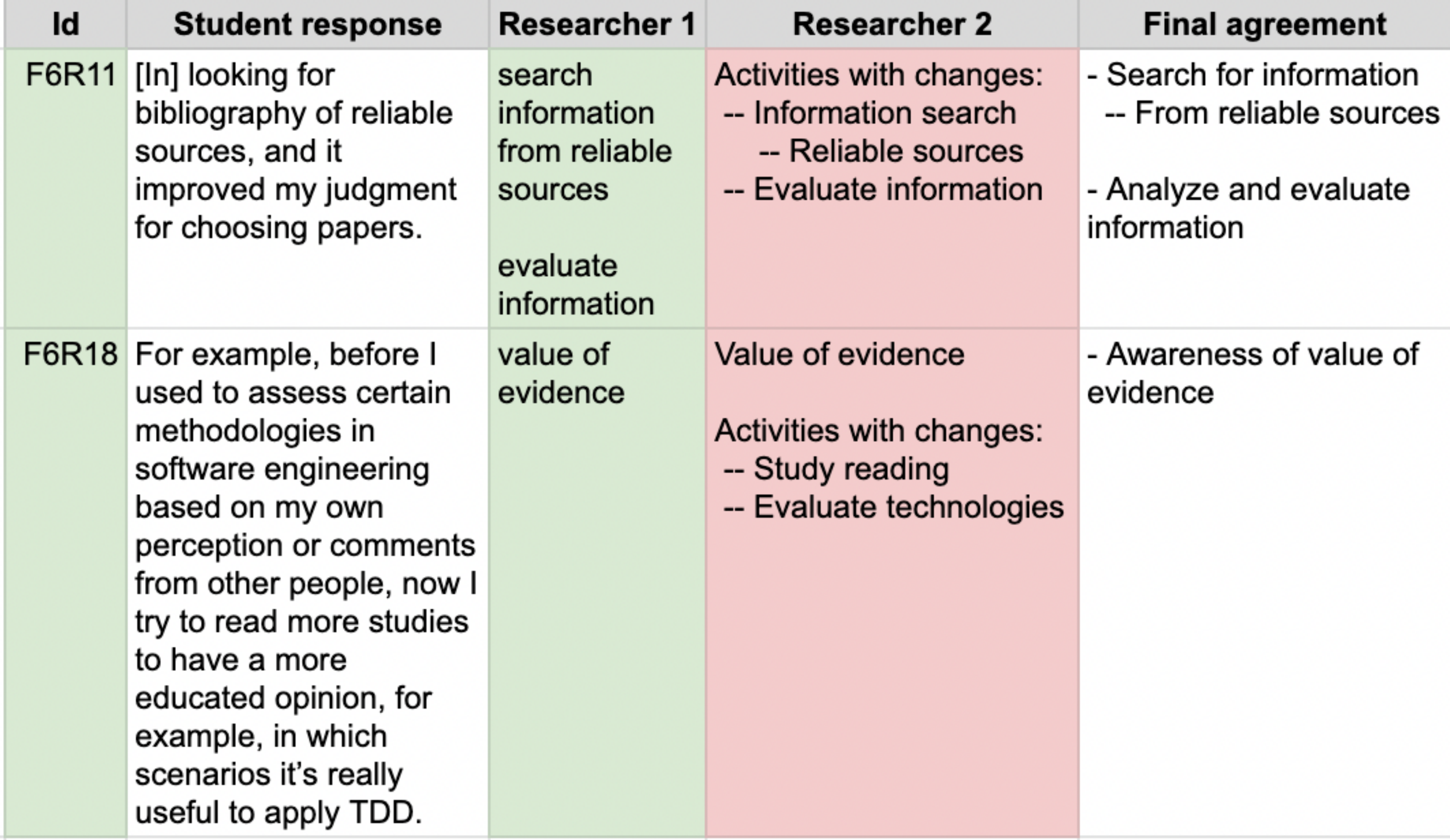}}
\caption{Fragment of coding for question F6 (`Changes motivated by knowing EBSE').} \label{fig2}
\end{figure}

As an example of the qualitative analysis carried out, Figure \ref{fig2} presents the coding of some F6 responses, including the codes identified by each researcher and the result of the agreement meeting.

\subsection{Second follow-up opinion survey}

Finally, twenty-one months after finishing the last course we conducted a second follow-up survey. With this survey, we investigated whether the students had used their training in professional practice (rather than an academic context) identifying which skills had been used and how often specific skills were mentioned.

Again, we designed a self-administered questionnaire with a set of closed and open questions. The questionnaire was designed in Google Forms and was available from August 27 to September 3, 2021. The translation of the questions (originally in Spanish) and the flow of the survey are shown below.

\begin{framed}
\small
\noindent
\begin{itemize}
    \item[S1] Are you working or have you worked in the software industry?
    \begin{itemize}
        \item Yes / No
    \end{itemize}
    \item[S2] Have you found useful in your professional (NOT academic) practice what you learned in the EBSE and SRs courses?
    \begin{itemize}
        \item Yes / No
    \end{itemize}
    \item[S3] \textit{[If answer to S2 was 'Yes']} Indicate what news skills from the course you used or found useful. If you can, add examples.
    \item[S4] \textit{[If answer to S2 was 'Yes']} On a scale from 0 (not at all) to 10 (all the time), how would you rate the frequency in which you use or have used things learned in the course in your professional practice?
    \begin{itemize}
        \item 0 / .. / 10
    \end{itemize}
\end{itemize}
\end{framed}

We counted the answers to each category of the closed questions. The textual responses to the open questions were collated and we used descriptive coding and the constant comparative method to classify the responses. 

Initially, \textanon{Pizard}{the first author} performed all of the analysis described above. Subsequently, \textanon{Vallespir}{the second author} reviewed the results. In a meeting, differences were discussed until an agreement was reached. Finally, \textanon{Pizard}{the first author} did the final aggregation of codes and the description of the results.

\section{Results}\label{sec_results}

\begin{figure}[htb!]
\includegraphics[width=8.7cm, left]{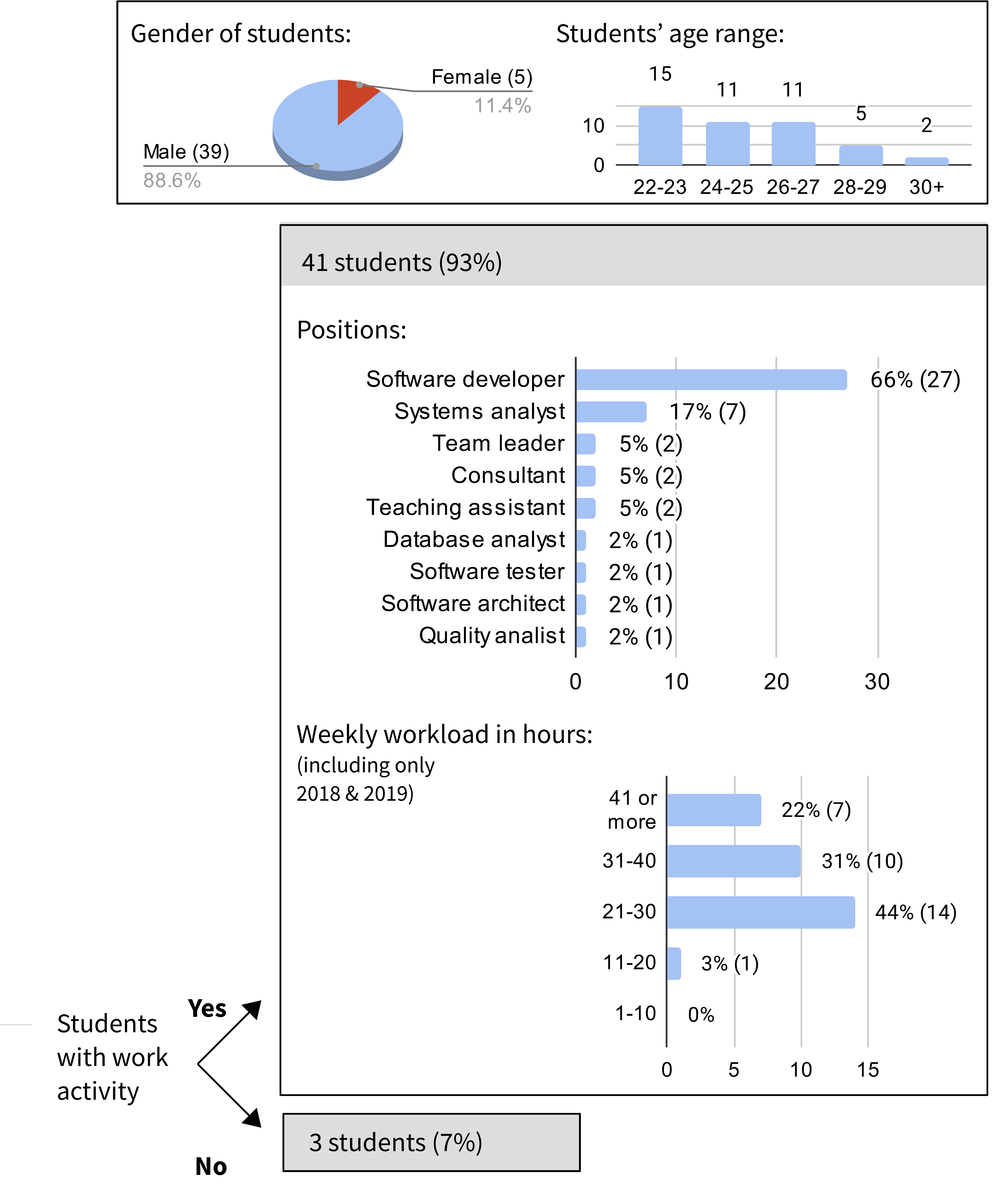}
\caption{Students demographics} \label{fig9}
\end{figure}

Figure \ref{fig9} shows demographic information describing our students obtained from the end of course survey. In particular, Figure \ref{fig9} reports the percentage of male and female students, their age range, and details of any paid employment. Three students reported having more than one position (e.g., team leader \&\ software developer), which we assume means they take on different roles in different projects. The vast majority of the students had worked in industry, and most of the positions that they held were technical rather than managerial. Our university is public and free of charge, and there is a lot of demand for IT workers. This is why most of our undergraduate students have part-time or full-time jobs in their last two years of studies.

\subsection{RQ1: Is our EBSE course adequate to train undergraduate students? }

The evaluation results for each course and for the first follow-up survey are summarized below.

\paragraph{Courses evaluations}

The three courses were taught one per year, from 2017 to 2019. Students' attendance was 10, 18, and 16 respectively. All students passed the module.

To assess the adequacy of our course we investigated (1) the levels of achievement of the learning outcomes, both perceived by the students and those assessed by the teachers, and (2) the students' opinions of the course, based on the survey and the final focus group.

First, we studied students' opinions and teachers' assessments of the achievement of learning outcomes, which were initially rated using a five-point assessment scale. Assessments allocated to scale points 3, 4 and 5 indicated that successful understanding and use of a topic was achieved\footnote{The five-point agreement scale used was: 1. Not achieved at all | 2. Very little achieved | 3. Satisfactorily achieved | 4. Almost completely achieved | 5. Completely achieved}. The subject success rate for a topic was calculated as the percentage of subjects who judged their own personal achievement to be successful for that topic. The team success rate was calculated as the percentage of team projects that were judged as successful for a specific topic by the course teachers. The left section of Table~\ref{table_courses_evaluation} shows the student-assessed success rates and the right section shows the teacher-assessed success rates for the team project. Topics that were not evaluated in the team assignments were left blank.

\begin{table}[]
\begin{threeparttable}
\caption{Student self-assessments of their learning achievements and teachers assessments team-based learning achievements.}
\label{table_courses_evaluation}
\small
\begin{tabular}{p{2.5cm}|ccc|ccc}
\toprule
                                         & \multicolumn{3}{p{2.2cm}}{Student self-assessed success rate} & \multicolumn{3}{|p{2cm}}{Teacher-assessed project success rate } \\  
\multicolumn{1}{c|}{Topic}                & 2019             & 2018             & 2017            & 2019           & 2018          & 2017          \\ \midrule
Basic aspects of scientific publications & 96\%             & 91\%             & 97\%            & -              & -             & -             \\
Evidence-based paradigm                  & 90\%             & 96\%             & 95\%            & -              & -             & -             \\
SRLs in SE                               & 100\%            & 100\%            & 100\%           & -              & -             & -             \\
Planning                                 & 97\%             & 100\%            & 95\%            & 94\%           & 81\%          & 92\%          \\
Search                                   & 93\%             & 94\%             & 98\%            & 100\%          & 100\%         & 100\%         \\
Selection                                & 84\%             & 86\%             & 92\%            & 92\%           & 93\%          & 100\%         \\
Quality assessment                       & 88\%             & 97\%             & 100\%           & 83\%           & 64\%          & 100\%         \\
Extraction                               & 98\%             & 100\%            & 100\%           & 100\%          & 100\%         & 100\%         \\
Mapping study analysis                   & 94\%             & 93\%             & 97\%            & 83\%           & 86\%          & 75\%          \\
Synthesis                                & 93\%             & 89\%             & 94\%            & 100\%          & 71\%          & 100\%         \\
Report                                   & 100\%            & 100\%            & 100\%           & 100\%          & 86\%          & 100\%         \\
Knowledge translation                       & 65\%             & 80\%             & 80\%            & -              & -             & -             \\
SR process                               & 100\%            & 100\%            & 100\%           & 100\%          & 100\%         & 100\%         \\
SE Research                              & 95\%             & 94\%             & -               & -              & -             & -            \\ \bottomrule
\end{tabular}
\begin{tablenotes}
    \footnotesize  
    \item The number of observations corresponds to the following detail. In 2017, 10 students \&\ 4 teams. In 2018, 18 students \&\ 7 teams. In 2019, 16 students \&\ 6 teams.
\end{tablenotes}
\end{threeparttable}
\end{table}

Looking at the self-assessment results, the learning achievement levels of all the topics seem quite good, since success levels are 65\% or above in all years. Some topics seemed consistently difficult to learn. Study selection, qualitative synthesis, and knowledge translation and diffusion were the three topics that some students in each year found relatively difficult.  With respect Knowledge translation, a limitation of the course was that it did not include any practical examples. For example, one student pointed out \textit{`I consider that I couldn't adequately address these objectives [those related to Knowledge translation] because, since there is no practical instances [in the course], and only remain theoretical, I couldn't understand it well'}.  For the other two topics, we did not find specific causes that explain why success was low nor why succes rates varied over the years, and we assume it is a natural result of dealing with different courses and different cohorts of students.

In the teamwork assessment, both planning and  mapping study analysis caused problems for a minority of teams in all three years. 

Thus, overall, learning achievement levels were mostly satisfactory, and problem topics were relatively consistent.

Students from all three courses expressed general satisfaction with the training. In all the focus groups held at the end of each course, the students expressed a positive opinion about the weekly work dynamics, i.e., the introduction of theoretical concepts and practical work in teams. They also valued positively the role of a practical assignment within the general course marking process.

\paragraph{Post-use evaluation}

In the first follow-up survey, seven months after finishing the last course, we asked the students if they had used lessons learned from the course and if the training received had been adequate for that use (see first survey, question F3).

All students who reported using EBSE (89\% of the total) indicated that the course trained them adequately. Five students added comments on the adequacy of the course. Four students confirmed in their comments that the course was complete. Although two added that it was complete considering it was a short course and another that it seemed complete given their superficial use of EBSE. Finally, another student indicated that the workshop approach and teamwork were aspects that greatly helped learning.

\subsection{RQ2: Does our EBSE course have any effects on the working practices of the students?}

The effects of EBSE training on students' attitudes and behaviors were studied in three instances: in the final survey of each course and in the two follow-up surveys. The results obtained are presented following the same sequence.

\paragraph{Perceived benefits of the course}

\begin{figure*}[htb!]
\includegraphics[width=12cm, left]{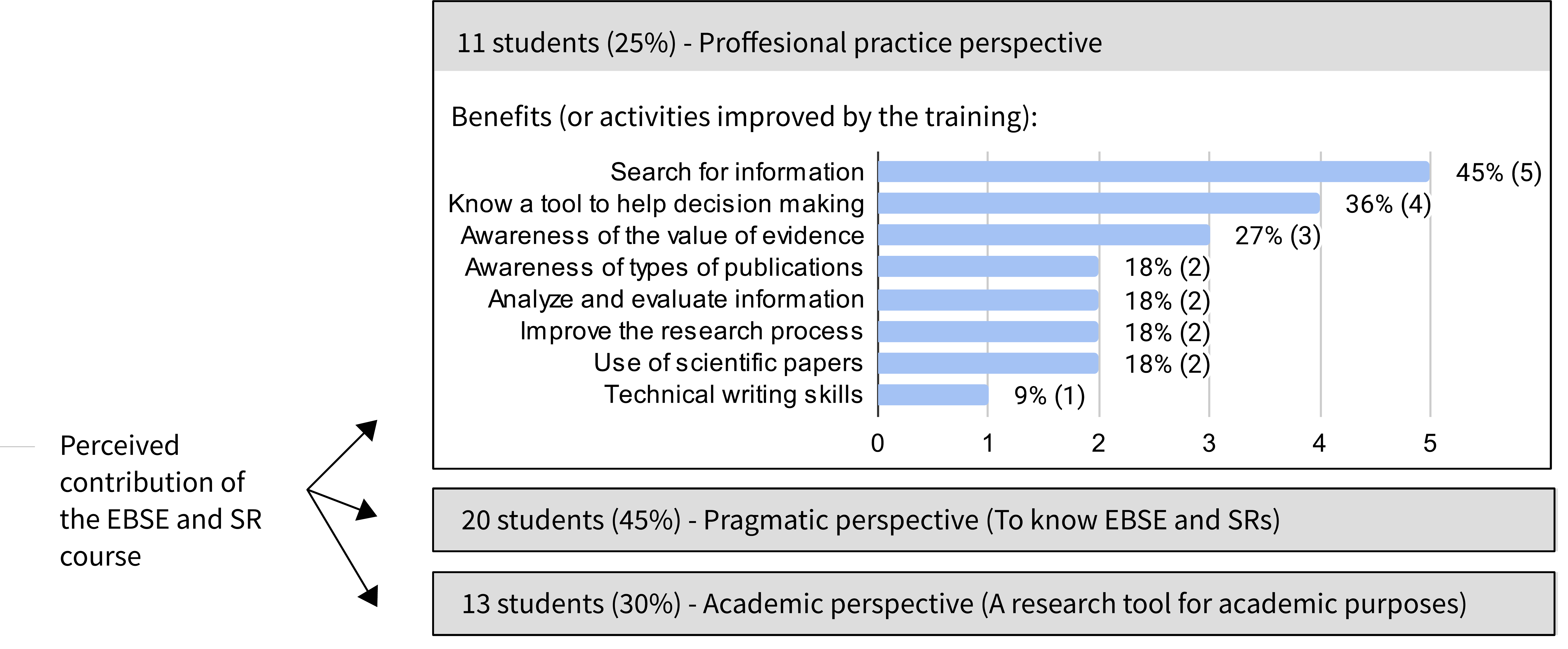}
\caption{Perceived benefits of the course (Nov 2017, Nov 2018 \&\ Nov 2019). } \label{fig7}
\end{figure*}

The students' opinion about the benefits of training\footnote{The question asked to the students was: \textit{`[SU-C1] Indicate, according to your criteria, what are the main contributions of the course to your education'.}} is grouped into three perspectives. Figure \ref{fig7} summarize their opinions. It also includes the analysis of the responses that indicated benefits in the work practices, since they are the results of interest to answer our research question.

First, 45\% of the students had, what we considered is, a pragmatic view and felt that the course helped them understand EBSE or how to conduct an SR. A student summarized the training benefits as follows: \textit{`Becoming aware of a systematic method, i.e., one that includes steps and procedures that were reviewed by experts in order to search and synthesize material that can answer questions'}.

Second, 30\% of the students reported  that after taking the course they have a tool to carry out more reliable research in software engineering. Two students thought that they did not see it as applicable in the industry, one of them explained: \textit{`Although we learned an interesting method for research in software engineering, for now I do not see it as very applicable in my working life'}.

Finally, 25\% believe that the course prepared them for better professional practice. In this regard, a student said: \textit{`On the one hand, [the course allowed me] to become more aware of all the baseless decisions that are made in our field. On the other hand, [I achieved] a greater knowledge about scientific studies and the different search engines'.}  The most perceived benefits were: improving information searching skills, having a new tool to support decision-making, and being aware of the value of the evidence. 

\paragraph{First follow-up survey}

\begin{figure*}[htb!]
\centering
\includegraphics[width=17.5cm]{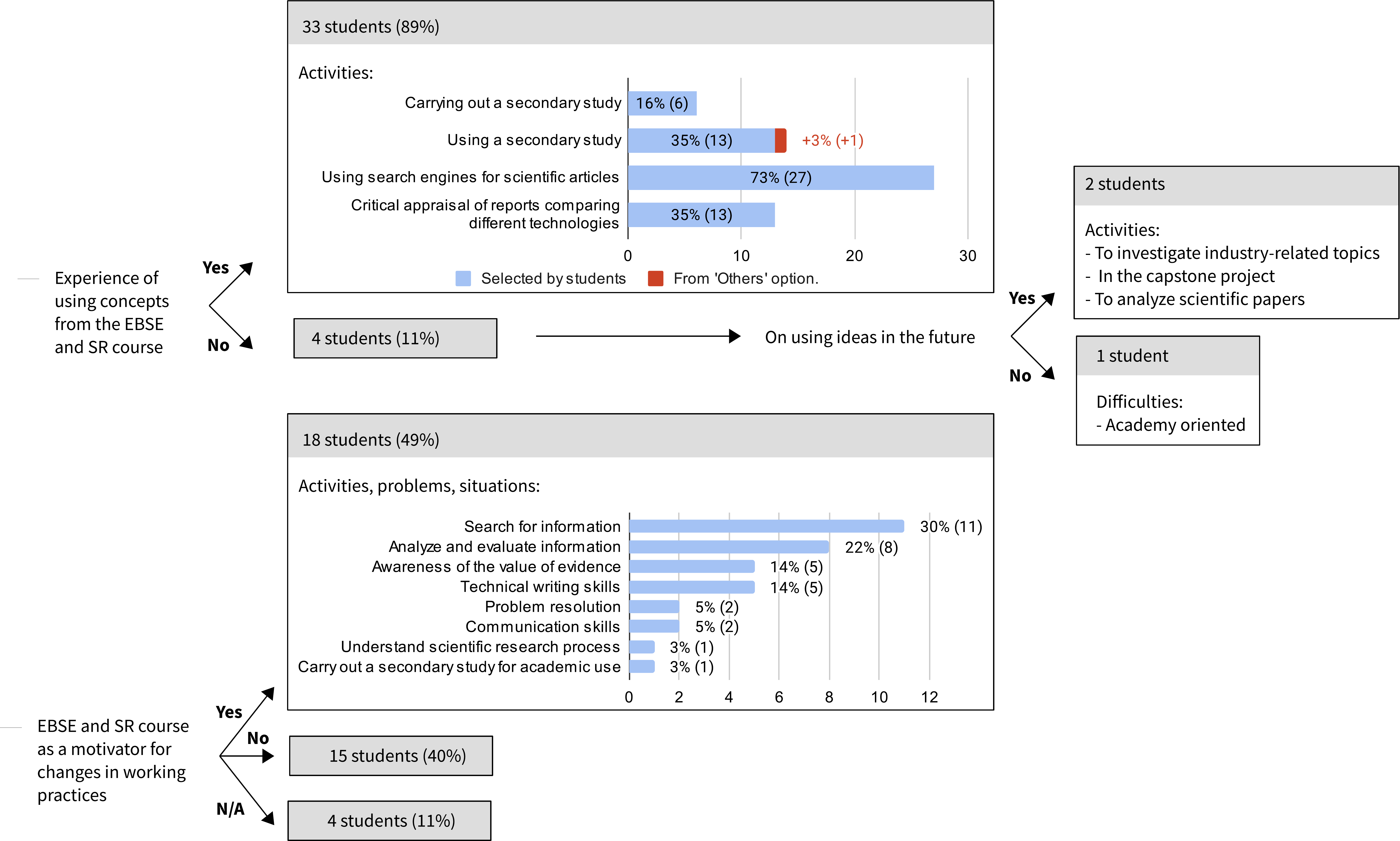}
\caption{Effects of the course on students' working practices (June 2020).} \label{fig4}
\end{figure*}

The survey was answered by 37 students, out of a total of 44 who took one of the courses (i.e., a response rate of 84\%).  Figure \ref{fig4} summarizes the results related to the impact of the EBSE course on the former students' behavior. To investigate this impact we used three students' views: experiences of using concepts learned in the course, their perspectives of using concepts in the future, and changes the course motivated to their working practices.

\textbf{EBSE usage experiences.} The majority of students (89\%) indicated having used concepts learned in the EBSE and SRs course (see the upper part of Figure \ref{fig4}).

Six students added answers in the free text option `Others'. One of them did not answer the question so it was discarded. Of the other five responses, all referred to the use of what was learned in the course (e.g., reading SRs or execution of some stages) in academia, that is, in their capstone project (4) or in other courses (1). We accumulated and highlighted in red in Figure \ref{fig4} any of these responses that semantically correspond to a category not previously selected by the student. 

The most frequently performed activity was using search engines to retrieve scientific articles, 73\% of those surveyed indicated having done this after completing the course. Using a secondary study and critically evaluating reports that compare technologies were the next two most performed activities (38\% and 35\% of respondents respectively). Lastly, conducting a secondary study was the least frequently performed activity, although it was done by 6 (16\%) of the students surveyed.

\textbf{Prospects of using EBSE concepts} Four students (11\%) indicated that they had not used concepts they learned in the course. Three of them responded about prospects for using ideas from the course in the future. One student expressed that a scientific approach is very valuable in decision-making but concludes that it is difficult to apply outside academic environments. The other two students thought that their EBSE knowledge could be useful in the future: to investigate industry-related topics, as a tool in the capstone project, and to analyze scientific papers).

\textbf{Changes motivated by knowing EBSE.} Almost half of the respondents (49\%) indicated that knowing EBSE motivated them to change their working practices. The activities in which the students noticed changes are shown in the lower part of Figure \ref{fig4}. 

For 30\% of the surveyed students, knowing EBSE motivated them to change the way they seek information. Some of them reported using EBSE to search for information for new or unusual problems, on topics with little information available or on topics about new technologies. Others said that the new knowledge allowed them to search more comprehensively or to use more reliable sources. One of them recognized changes in: \textit{`looking for bibliography of reliable sources, and it improved my judgment for choosing papers'}.

Students also identified information assessment and the value of the evidence as activities that were positively influenced by the course (22\% and 14\% respectively). Both of these skills are related and indicate enhancement in the students' abilities to evaluate the information they collect or receive. One student said: \textit{`For example, before I used to assess certain methodologies in software engineering based on my own perception or comments from other people, now I try to read more studies to have a more educated opinion, for example, in which scenarios it’s really useful to apply TDD'}.

Some students also indicated improvements in their technical writing skills (14\%), problem-solving skills (5\%), and communication skills (5\%). To illustrate the latter, some students reported being able to better communicate or substantiate their ideas. These improvements may not solely be due to the knowledge of EBSE, but may also be due to the methodology and activities of the course.

Finally, one student reported a better understanding of the scientific research process, and another to have carried out a secondary study in their academic activity.

\paragraph{Second follow-up survey}

The second survey was answered by 29 students (i.e., a response rate of 66\%). All indicated that they were working or had worked in the software industry (S1). 

\begin{figure*}[htb]
\includegraphics[width=15cm, left]{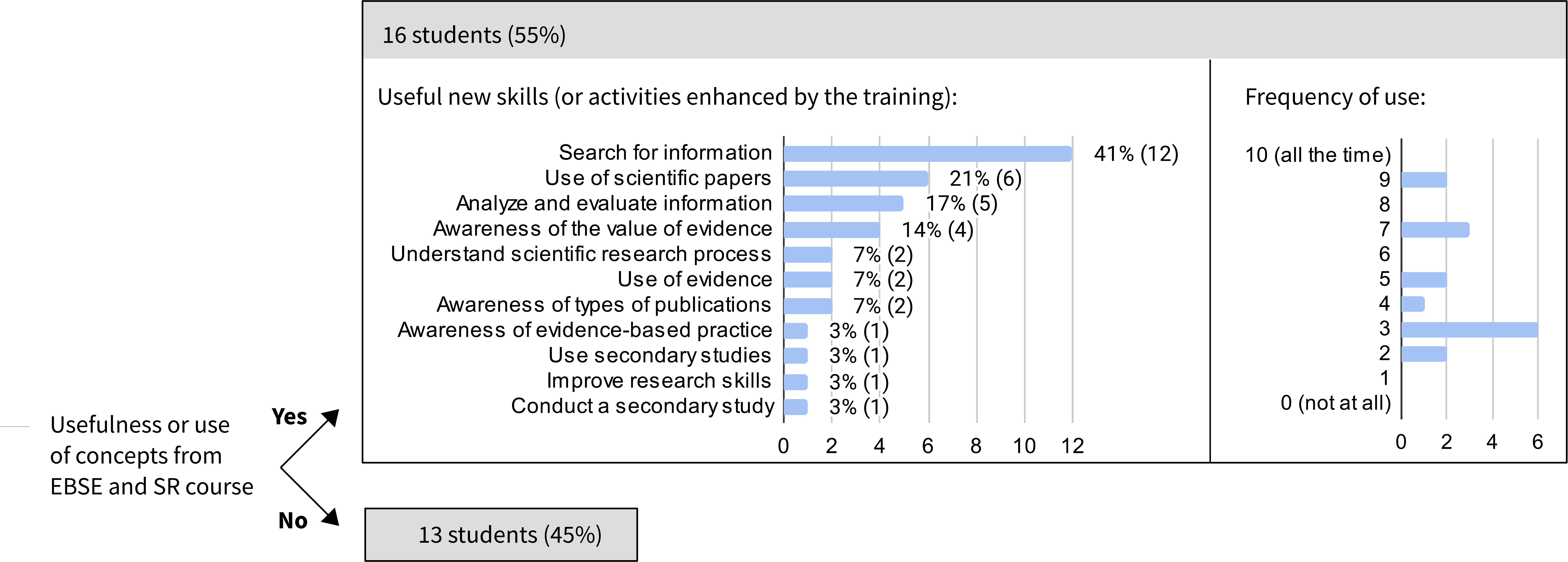}
\caption{Effects of the course on students' working practices (August 2021)} \label{fig5}
\end{figure*}

A 55\% of respondents found what they learned in the EBSE course useful in their (non-academic) professional practice (S2). Figure \ref{fig5} shows skills from the course that respondents indicated were useful or used (S3), and the frequency of use (S4). In S3, many students did not indicate skills learned in the course that they use (what was requested) but the activities improved by the training. Since both correspond to the effects of the training, our analysis of that response includes both useful new skills and activities enhanced by the training. One of the answers to S3 did not answer the question so it was discarded. 

Four out of ten of those surveyed said they have improved the way they search for information after taking the course. On this a student commented that now they can: \textit{`recognize useful information, from reliable sources and carry out more precise searches'}.

A fifth reported using scientific literature after taking the course. One former student said: \textit{`I learned to use scientific articles as a source of information and evidence on a day-to-day basis (and how to search for them correctly) […] It also helped me to look for reviews at times I wanted to know quickly the state of the art on a specific topic'}. Related to this, two students also reported understanding the scientific research process was a useful new skill.

As in the previous survey, some of the former students reported improvements in their abilities to assess and value the information, both in the evidence assessment and in the analysis and evaluation of information. Two students also indicated that they were more aware of the different types of publications. One of them claimed: \textit {`I also believe that the training from the course allows me to identify the different types of papers, whether they are primary studies, reviews, etc. and this is essential when looking for information'}.

Some students also reported using evidence. In our course, we do not teach concepts and techniques to address in detail the use of evidence and students may not know if they are actually using evidence. This is a non-trivial problem, for example, the term evidence is already controversial in other areas \citep{rycroft2004}.

Other impacts of the EBSE training were reported to be: awareness of the evidence-based practice, using secondary studies, improving research skills, and conducting secondary studies.

\subsection{Discussion of findings}\label{sub_sec_discussion}

Several aspects of our results deserve reflection and analysis.

\paragraph{Course adequacy} 

Our training was perceived as adequate by the teachers when assessing students' teamwork assignments and by the students themselves at the end of each course and after they tried to use what they learned in practice. In the latter case, all students who reported using what they were taught found the training adequate.

In health care and medicine, there is a greater formalization of the teaching of evidence-based practice (EBP), for example, there are proposals for EBP core skills (see for example \citep{albarqouni2018}) and recognized tests for the minimum skills required \citep{ramos2003}. These disciplines have stringent accreditation requirements because of their direct impact on human welfare. However, having such tools in software engineering could help us to better and more consistently assess the adequacy and the effects of EBSE training. Our training proposal is based on the concept of learning outcomes (see \citeanon{pizard2021} for the full description) that could be used as a starting point for discussions on this issue.

\paragraph{Effects of EBSE training} 

Although few students reported having read or participated in the conduction of secondary studies, their responses confirm that the training allowed them to learn and apply the principles of evidence-based practice. The most reported effects in both surveys were: improvements in information search and analysis skills, and awareness of the value of the evidence. In the second post-course survey, the use of scientific papers was also frequently mentioned. The students reported different frequencies of use of the EBSE-related skills, but overall usage appears to be more than sporadic.

In brief, our EBSE training provided the students with an awareness of research and scientific evidence and improved their information gathering and information literacy skills. All of these effects are consistent with previous research assessing the impact of EBP training. For example, several nursing studies report that EBP training improves students' search skills and information literacy skills \citep{horntvedt2018}, and also improves students' research skills and their confidence in research \citep{patelarou2020}.

The effects of our training are consistent with the initial opinions of the students about the benefits of the course. Also, they are consistent with the previous studies on the benefits of learning EBSE (i.e., improvements in research skills, awareness of the value of adding evidence, and improvement in the search and organization of information \citep{kitchenham2010, baldassarre2008}). This could indicate: (1) a relationship between initial attitudes to EBSE and subsequent use of  EBSE-related skills, and (2) that opportunities for individuals to reflect on the benefits of a training course could help them to identify ways to use EBSE related-skills at a later date. The first issue is under study in other areas of EBP. For example, it has been shown that knowledge of EBP and practitioners' attitudes towards EBP influence its adoption \citep{stroobants2016} and that attitudes towards EBP are considered key for its adoption \citep{wike2014}. The second issue requires further research, for example, by exploring reflective instances in groups and investigating whether early adopters can influence other students.

Our results seem to indicate, we believe for the first time, that EBSE training makes practitioners more confident in the value of scientific evidence and fosters its use for support decision-making. To our knowledge, this is also the first study that provides support for the potential of the proposals (i.e., \citep{devanbu2016, legoues2018}) to use EBSE to bring practitioners closer to scientific evidence and collaborate in closing the gap between academia and industry.
    
\paragraph{Future of EBSE education} 

Currently, curricula guides for undergraduate students in CS and SE do not consider evidence-based practice \citep{cscurricula2013, securricula2014}. Certainly, more studies are needed to justify the inclusion of EBSE, for example, in crowded curriculums common in Europe and the USA.

However, in other disciplines, EBP teaching is more strongly promoted given the benefits of its use by practitioners. For example, evidence-based practice has been included as a core component of the curriculum of undergraduate, postgraduate, and continuing education of health programs throughout the world, and many accreditation boards also expect all clinicians to have competent training in EBP \citep{albarqouni2018}.

It is also necessary to point out that, from our understanding, each discipline has its own tailored content and approaches to teaching evidence-based practice. This seems to be also the case in our field, although EBSE appears to be similarly teachable to SE and CS students. This study serves as a sample of this since in our university there is a unique curriculum (in particular, CS focused).

Given the continuous renewal of our discipline, it may be necessary to reimagine what skills tomorrow's software engineers will require and explore ways to bring their practice closer to scientific evidence. To our knowledge, there is a lack of studies on the benefits of adopting EBSE in non-academic environments. Even so, we believe it is important to begin to discuss the inclusion of EBSE teaching in curricula guides, since, as our results show, its teaching can contribute to its adoption and to the use of scientific evidence in professional practice. As Mary Shaw, in the first prospects on the teaching of SE, claimed: \textit{`The greatest danger to software engineering curriculum designers is lack of imagination'} \citep{shaw1987}. 

\subsection{Limitations of this study.} 

Although the results reported in this study have addressed the fact that our previous evaluation of the training module was based on a small sample, many of the limitations discussed in our previous paper \citeanon{pizard2021} apply equally to this study. In particular, the limitations related to the good subject problem among voluntary participants, using incentives to encourage participation at the end of each course, and the course emphasis on SRs rather than EBSE apply also to this study. 

Question F6 of the first follow-up survey caused some confusion. The question is a continuation of question F5 and it sought to investigate which activities of their working practices the students had improved from learning EBSE. Two responses out of a total of 18 included references to academic (and non-professional) activities. Many other answers did not indicate whether they correspond to professional or academic activities. Nonetheless, we did not discard data during analysis, because it confirms that EBSE training has general benefits to undergraduate students. In the second follow-up survey we specifically investigated benefits related to the students' industry-related working practices.

Surveys rely on respondents' reflection and reporting on their attitudes and behaviors, and this can bias the results in different ways \citep{lethbridge2005}. For example, people tend to remember events that are most important to them. Although we tried to mitigate its effects by asking participants several different questions about the same issue and requesting examples to support their previous answers, our results must be considered with this limitation in mind.

\section{Concluding remarks}\label{sec_conclu}

Our study reports that EBSE training has positive impacts on the professional practices of those trained. In particular, the majority of our EBSE course students reported using their training, not just in an academic context, but also in their jobs in industry. They reported a greater awareness of research and evidence, and improvements to their information search and analysis skills. As a consequence, EBSE training could be useful to train practitioners in concepts and techniques of evidence-based practice but also to foster collaboration between industry and academia.

The major novelties of our study are that it:

\begin{itemize}
    \item Involves the delivery of three courses of EBSE and SRs to university students based on a predefined set of learning outcomes.
    \item Is an attempt to evaluate the impact of an EBSE training on the attitudes and behaviors of the trainees (particularly, in their work practices) in different periods.
    \item Confirms that EBSE training has enabled more than half of students to enhance their work practices, by improving their information gathering and analysis skills, and through better awareness of evidence and research.
    \item Confirms that the benefits of teaching EBSE are similar to those obtained from teaching EBP in other disciplines.
    \item Suggests that EBSE training improves the performance of novice industry practitioners not just individuals with conventional decision-making roles such as senior software engineers or project and quality managers.
\end{itemize}

EBP training in medicine has been shown to be more effective if it is integrated or supported with clinical practice \citep{ilic2014, larsen2019}, also bringing practitioners even closer to scientific evidence. This could be achieved in EBSE training, for example, by teaching EBSE using scenarios of use of evidence created together with practitioners (as suggested by Manns and Darrah \citep{manns2012}). Knowing the decision-making process in the software industry seems to be the key to orienting EBSE training to the needs of the industry. To address this, it could be useful to survey managers of software companies asking them who makes decisions about SE methods and tools and how such decisions are made. It has been suggested that analyzing and making the decision-making process visible can improve EBP adoption \citep{benfield2021}. In addition, using theoretical proposals for studying the decision-making process (e.g., \citep{burge2008}) could also contribute in this direction.

\begin{acks}
First of all, we thank all the students who participated in our EBSE courses. We thank Fernando Acerenza and Cecilia Apa, of the School of Engineering, Universidad de la República, Uruguay, for their collaboration as teachers in the 2017-2018 and 2019 EBSE courses, respectively. We also thank Ximena Otegui, of the Teaching Department, School of Engineering, Universidad de la República, Uruguay, for her collaboration in creating and improving the LOs of the course. Finally, we thank the reviewers for their helpful comments, which have allowed us to improve our manuscript.
\end{acks}

\bibliographystyle{ACM-Reference-Format}
\bibliography{main}

\end{document}